\documentstyle[prl,preprint,aps]{revtex}

\def\Dslash{\raise.15ex\hbox{/}\kern-.77em D}

\draft
\begin{document}
\preprint{\vbox{\hbox{SNUTP 94-134} \hbox{hep-ph/9505420}}}

\title{\bf Light-Front View of The Axial Anomaly}

\author{ Chueng-Ryong Ji$^{1,3}$ and Soo-Jong Rey$^2$}
\address{ Department of Physics, North Carolina State University \\
          Raleigh, North Carolina 27695-8202 USA$^1$ \\
          Department of Physics \& Center for Theoretical Physics\\
          Seoul National University, Seoul 151-742, Korea$^2$\\
          Institute for Nuclear Theory, University of Washington\\
          Seattle, Washington 98195$^3$} 

\date{\today}
\maketitle
\begin{abstract}
Motivated by an apparent puzzle of the light-front vacua incompatible
with the axial anomaly, we have considered the two-dimensional massless 
Schwinger model for an arbitrary interpolating angle of the quantization 
surface. By examining spectral deformation of the Dirac sea under an 
external electric field semiclassically, we have found that the axial 
anomaly is quantization angle independent. This indicates an intricate 
nontrivial vacuum structure present even in the light-front limit. 
\end{abstract}

Recently the idea of using light-front quantization \cite{dirac}, which 
has been applied successfully in the context of current algebra \cite{lcca} 
and parton model \cite{lcparton} in the past, was 
revived as a promising method for solving QCD \cite{brodsky,wilson}.
While the hope is partly based on the observation that the perturbative 
vacuum becomes extremely simple, there has never been any serious study of 
nontrivial vacuum states such as chiral symmetry breaking and $\theta-$vacuum 
structures. As these aspects are essential to the low-energy hadron 
physics, it is important to understand how these aspects come about in the 
light-front quantized QCD.  In simpler models, this issue has been studied 
only very recently \cite{zeromode}, and found that the $k^+ = 0$ 
zero-modes are responsible for nontrivial vacuum phenomena.

In this Letter, we address another particular aspect of the nontrivial vacuum 
structure: the axial anomaly \cite{anomaly1}. 
It is well-known that regularization procedure 
in quantum field theory sometimes cannot preserve all of the classical 
symmetries. The axial anomaly represents a paramount example of such phenomena.
The axial anomaly may be understood as an ultraviolet phenomena stemming
from the linear divergence of the one-loop diagrams. Since the diagram is 
divergent, one has to regularize, for example, through Pauli-Villar method.
It is also known that the axial anomaly can be understood as an infrared 
phenomenon arising from the level crossing of the Dirac vacuum in the 
background of nontrivial gauge fields. The level crossing is nothing but 
a pair creation of the charged fermions. This interpretation is apparently 
incompatible with the light-front quantization since the fermion pair 
creation is not possible due to the $k^+$ momentum conservations. It is 
therefore the purpose of this letter to see how the light-front quantized 
fermionic vacuum responds to produce the axial anomaly phenomena. For 
simplicity, we will only consider the two-dimensional Schwinger model 
\cite{schwinger}. 

We study the massless Schwinger model quantized with an 
interpolating coordinates \cite{kent} $(x^+,x^-)$, 
\begin{equation}
{x^+ \choose x^-} \equiv { \sin{\theta/2} \hskip0.7cm \cos{\theta/2}
\choose \hskip0.1cm \cos{\theta/2} \hskip0.3cm - \sin{\theta/2}} 
{t \choose x}.
\label{coordinates}
\end{equation}
The $x^+, x^-$ are taken as time and space coordinates respectively.
These coordinates interpolate between the equal-time one at $\theta = \pi$ 
and the light-front one at $\theta = \pi/2$. For an arbitrary interpolating
angle $\theta$ we denote $c \equiv - \cos \theta$ and $s \equiv \sin \theta$.
In what follows, we put the space coordinate $x^-$ on a compact circle of 
circumference $L$ to regulate the infrared limit. The action of the Schwinger 
model is given by
\begin{equation}
{\cal S} = \int dx^+ dx^- [-{1 \over 4 } F_{\mu \nu} F ^{\mu \nu} + 
\bar \psi i \gamma^\mu (\partial_\mu + i e A_\mu) \psi].
\label{eq:fermi}
\end{equation}
Here, $\psi$ denotes the two-component Dirac spinor and 
$\bar \psi \equiv \psi^\dagger \gamma^0$.
We choose the Dirac matrices exclusively in the chiral representation.
At an arbitrary interpolating angle the Dirac matrices in the chiral
representation are given by 
\begin{equation}
\gamma^\pm = {\hskip0.9cm 0 \hskip0.6cm \pm \sqrt {1 \pm s} \choose 
\sqrt {1 \mp s} \hskip1cm 0 \hskip0.4cm}.
\label{gmatrix}
\end{equation}
They satisfy Dirac matrix algebra 
$\{ \gamma^\pm, \gamma^\pm \} = 2 g^{\pm\pm} {\bf I} = \pm 2 c {\bf I}$ 
and 
$\{ \gamma^+, \gamma^- \} = 2 g^{+-} {\bf I} = 2 s {\bf I}$.
At equal-time limit $c \rightarrow 1$ and $s \rightarrow 0$, 
$\gamma^\pm$ become $\gamma^0 = \sigma^1$ and 
$\gamma^1 = i \sigma^2$
respectively. Note that the (anti)-Hermiticity property $\gamma^{0 \dagger} = 
+ \gamma^0$, $\gamma^{1 \dagger} = - \gamma^1$ is valid only at equal-time 
limit. For a generic interpolating angle $\gamma^\pm$ are neither Hermitian nor 
anti-Hermitian; they are related each other through $\gamma_5 \equiv \gamma_o
\gamma_1 = \sigma_3$ as $\gamma^{\pm \dagger} = \mp \gamma^\mp \gamma_5$. 
More generally $\gamma_5$ defined as $\gamma_5 \equiv (\gamma^+ \gamma^- - 
\gamma^- \gamma^+)/2 = \sigma_3$, independent of the interpolating angle 
$\theta$. Therefore it is always possible to decompose the 
Dirac fermion into LH and RH components for an arbitrary angle $\theta$

%% In fact, this is not clear at all.
%% If I make a connection, then I find that you set 
%% (b_n e^{-i P_n x^-} - d_n^\dagger e^{+i P_n x^-})/{\sqrt 2}
%% as a_{Ln} e^{+i P_n x^-} in the upper component of the spinor
%% and 
%% (b_n e^{-i P_n x^-} + d_n^\dagger e^{+i P_n x^-})/{\sqrt 2}
%% as a_{Rn} e^{+i P_n x^-} for the lower component.
%% However, note that the upper component has the multiplication
%% factor \psi_{Ln}=\theta(-n) and likewise the lower component 
%% is multiplied by \psi_{Rn}=\theta(+n).
%% Thus, the upper component has only one sign in n such as
%% n= -1/2, -3/2, -5/2, etc and, since P_n = {2 \pi n}/L,
%% Due to this constraint,
%% I cannot reduce my relations as you presented above.
%% Of course, the lower component has the same problem.
%% Thus, the representation of the spinor as the way
%% presented in our paper seems to me correct only 
%% modulo the e^{\pm{i} P_n x^-} factors.
%% Therefore, the Eq.(4) seems to me not exactly correct
%% but only schematically correct.
%% I wouldn't mind to keep the equation since it is 
%% at least schematically correct, but these constraints 
%% must be understood to be clear.  

\begin{equation}
\psi = {\psi_L \choose \psi_R} = {1 \over \sqrt L} \sum_{n}
{ \psi_{Ln} {\bf a}_{Ln} \choose \psi_{Rn} {\bf a}_{Rn}} e^{i p_n x^-}
; \hskip1cm p_n = 2 \pi n / L. 
\label{modeexp}
\end{equation}
The fermions satisfy an anti-periodic boundary condition $\psi(x^- + L) = 
- \psi (x^-)$. Hence, the momentum quantum number is half-integer valued 
$n \in {\bf Z} + 1/2 $. 
Free fermions of each chirality satisfies the following 
equations of motion
\begin{eqnarray}
&[& c \, \partial_+ + (1 + s) \, \partial_- ] \psi_L (x^+, x^-) = 0,
\nonumber \\
&[&(1 + s) \, \partial_+  - c \, \partial_- ] \psi_R (x^+, x^-) = 0.
\label{eqns}
\end{eqnarray}
Solving the above equations of motion,
the basis wave functions $\psi_{Ln}, \psi_{Rn}$ in Eq.(4) 
are given by
\begin{eqnarray}
\psi_{Ln} &=&  { \theta (-n) } ( 1 - s)^{-1/4},
\nonumber \\
\psi_{Rn} &=& {\theta (+n) } (1+s)^{-1/4}.
\end{eqnarray}

The canonical quantization then proceeds with an anticommutation relation 
$ \{ \pi_\psi(x) , \psi(y) \}_{x^+ = y^+} = i \delta (x^- \, - \, y^-)$
where $ \pi_\psi(x) = \partial {\cal L} / \partial (\partial_+ \psi(x))
= i \overline \psi(x) \gamma^+$ .
In terms of chiral fermion components, 
$\sqrt {1- s} \{ \psi^\dagger_L(x), \, \psi_L(y) \}_{x^+ = y^+} =
\sqrt {1+ s} \{ \psi_R^\dagger , \,  \psi_R (y) \}_{x^+ = y^+} =
\delta (x^- \, - \, y^-). $
Consequently
the mode creation and annihilation operators satisfy
$ \{{\bf a}_{Ln}, {\bf a^\dagger }_{Lm} \} = 
\{ {\bf a}_{Rn}, {\bf a^\dagger }_{Rm} \} = \delta_{n, m}
$ and $ \{ {\bf a}_{Ln}, {\bf a}_{Rm} \} = 0. $
For the gauge field we choose a gauge $\partial_- A_- = 0$ gauge. 
Not only permitting a 
manifestly periodic boundary condition for $A_+$, this gauge choice
has an advantage of representing a constant electric field as 
$\partial_+ A_-^{(o)}(x^+)$ where $A_-^{(o)}(x^+)$ is the $x^-$ independent, 
zero mode of $A_-$. 
The canonical momentum conjugate to $A_-^{(o)}$, 
$\Pi_o = \int dx^- \, \partial {\cal L} / \partial (\partial_+ A_-^{(o)}) = 
L \partial_+ A_-^{(o)}$ satisfies the usual commutation relation 
$ \Big[ \Pi_0 (x^+), A_-^{(o)}(x^+)] = - i$.

Hamiltonian of the Schwinger model projected to gauge zero-mode is 
expressed as
\begin{eqnarray}
P_+^{(o)} &=& \Pi_0 \partial_+ A_-^{(o)} 
+ \int d x^- \, \Big(\,  \pi_\psi \partial_+ \psi  - {\cal L}  \, \Big)
\nonumber \\
&=& {L \over 2} E^2 
+ \int dx^- \, \Big[ - \sqrt {1 + s} \, \psi_L^\dagger \, i \partial_- 
\psi_L + \sqrt {1 -s} \psi_R^\dagger \, i \partial_- \psi_R
+ e A_-^{(o)} J^- + e A_+ J^+ \, \Big].
\label{hamilton}
\end{eqnarray}
The gauge currents $J^\pm$ are related to the chiral currents 
$J_{L,R} \equiv \sqrt {1 \mp s} \psi^\dagger_{L,R} \, \psi_{L,R}$ as
$J^\pm = \sqrt {1 \mp s} \, \psi^\dagger_L \psi_L \, \pm \, \sqrt {1 \pm s}
\, \psi_R^\dagger \psi_R
$
so that
$
J^+ = J_L + J_R, J^- = (1+s) / c \,  J_L - (1 - s) / c \, J_R.$
As it stands, however, $J^\pm$ are ill-defined because of short-distance 
singularity of composite operators. One may define them in terms of the 
currents regularized by Schwinger's point-splitting method
along the $x^-$ space direction
\begin{equation}
J^\pm [\epsilon]  \equiv \overline \psi(x^- + \epsilon) \, \gamma^\pm \,
e^{-i \, e \int_{x^-}^{x^-+\epsilon} d x^- \, A_-^{(0)} (x^+) }\, \psi (x^-)
= \overline \psi (x^- + \epsilon) \, \gamma^\pm \,
e^{- i \, e \, \epsilon A_-^{(o)}(x^+)} \, \psi (x^-).
\end{equation}
Using the short-distance behavior of the fermion bilinear 
$\psi^\dagger_{L,R} \psi_{L,R} \rightarrow \pm i / 2 \pi \epsilon 
\sqrt {1 \mp s}$, it is straightforward to evaluate the regularized 
currents as
\begin{equation}
J^+ = :\! J^+\! :,  \hskip1cm J^- = :\!J^-\!: + {e \over \pi c} A_-^{(o)}.
\end{equation}
Thus, gauge invariant regularization leads to an important modification of 
the Hamiltonian $P_+^{(o)}$ for the quantum mechanics of $A_-^{(o)}$:
\begin{equation}
P_+^{(o)} = 
{1 \over 2 L} \Pi^2_{0} + e A_-^{(o)} \, \Big({s \over c} :\!Q\!: 
+ {1 \over c} :\!Q_5\!: \Big)
+ {e^2 \over \pi c} L (A_-^{(o)})^2
\label{zerohamilton}
\end{equation}
in which $Q_{L,R} \equiv \int \! dx^- \, J_{L,R}$ and 
$Q = Q_L + Q_R, \, \, Q_5 = Q_L - Q_R$.
The Heisenberg equation of motion for the zero-mode $A_-^{(o)}$ then reads
\begin{equation}
L \partial_+^2 A_-^{(o)} = 
- e \, \Big({s \over c} :\!Q\!: + {1 \over c} :\!Q_5\!:\Big) - 
{2e^2 \over \pi c} L A_-^{(o)}.
\end{equation}

Because of the gauge invariance $A_+ \rightarrow A_+ + \partial_+ \Lambda$
the electric charge is manifestly conserved $d Q / d x^+ = 0$. On the other
hand, for a constant electric field background  $E = - \partial_+  A_-^{(o)}$,
$x^+$-derivative of the equation of motion for $A_-^{(o)}$ becomes
\begin{equation}
\partial_+ Q_5 = {2 e \over \pi} L \, E
\label{splitanom}
\end{equation}
viz. the axial charge is anomalously produced at a rate proportional to 
the constant electric field $E$. This is precisely the axial anomaly in
the massless Schwinger model.

In the above method, the axial anomaly has arisen from regularizing 
short-distance singularities of coincident quantum operators. As such, the
axial anomaly may be interpreted as a ultraviolet phenomena. 
Alternative method is a direct calculation of the relevant Feynman diagrams. 
At the light-front limit, the axial anomaly was calculated in this 
way \cite{jones}. 
However the calculation was rather involved compared to the perturbative 
calculation in the equal-time limit. 
On the other hand it is well-known that the axial anomaly may be understood 
in yet another, semiclassical way \cite{anomaly} through spectral 
flow \cite{spflow} crossing the zero-energy level in the Dirac vacuum
and simultaneous pair production of left- and right-moving fermions under 
appropriate external gauge field. This alternative interpretation 
emphasizes that the nontrivial structure of the Dirac vacuum and that the 
axial anomaly is rather an infrared phenomenon. However, this interpretation
of the axial anomaly poses a serious interpretational problem at a first 
glance since Eq.(5) indicates that half of the fermion degrees of freedom
decouple at the light-front limit $c = 0$. It is not clear at all how the 
anomaly is understood as a pair production of fermion pairs of opposite 
chirality. 

In order to resolve this difficulty, we first construct the Dirac vacuum at 
an arbitrary interpolating angle $\theta$. We define the Dirac vacuum by
filling all the negative energy levels of the left- and right-moving fermions
to the Fock vacuum:
\begin{equation}
|{ \rm vac.} \bigr> \equiv \prod_{\{n \le -1/2  \} } \,
{\bf a}^\dagger_{L, +n} \, {\bf a}^\dagger_{R, -n} \, | 0 \bigr>.
\end{equation}
As is clear the Dirac vacuum is not evenly populated between the left and the
right moving chiral modes of fermion. This leads to an important consequence
for correctly accounting for the axial anomaly and the vacuum energy.

Chiral charge operators are defined by Eqs.(12, 13). In terms of mode operators
they are given by
\begin{eqnarray}
Q_L &\equiv& \int dx^- J_L :=: 
\sum_{n,LH} {\bf a}_{Ln}^\dagger \, {\bf a}_{Ln} ,
\nonumber \\
Q_R &\equiv& \int dx^- J_R :=: 
\sum_{n,RH } {\bf a}_{Rn}^\dagger \, {\bf a}_{Rn}.
\label{charges}
\end{eqnarray}
Let us calculate vacuum chiral charges in an external electric field by
evaluating expectation values of $Q_{L,R}$ over the Dirac vacuum. 
As they stand, the vacuum chiral charges are formally infinite, hence, 
ill-defined. We therefore define regularized chiral charges by cutting off the 
ultraviolet negative modes. In order to preserve the gauge invariance, the 
regularization procedure has to preserve the gauge invariance.
Thus 
\begin{eqnarray}
\Big< Q_L({\rm reg.}) \Bigr> &=& \sum_{n \le -1/2} 1 \cdot 
\exp \Big[\,  \epsilon {1+s \over c} \Big(+p_n - e A_-^{(o)}(x^+) \Big) 
\, \Big],
\nonumber \\
\Bigl< Q_R ({\rm reg.}) \Bigr> &=& \sum_{n \ge + 1/2} 1 \cdot 
\exp \Big[\, \epsilon {c \over 1+s} \Big(- p_n + e A_-^{(o)}(x^+) \Big) \,\Big].
\label{regcharges}
\end{eqnarray}
Here, $\epsilon$ is the regularization parameter and $p_n = {2 \pi n}/L$ for 
half-integer-valued $n$. 
Physical quantities will be defined by a finite contribution of the 
regularized quantities as the regulator is removed $\epsilon \rightarrow 0$.
Note that we have introduced the exponential regularization factor which 
depends explicitly on the interpolating angle $\theta$. This means we need
to regularize in an asymmetric manner between the left- and the right-moving
fermions. We will see in the following why this is the correct regularization 
scheme. For now we note that, at equal-time limit $\theta \rightarrow \pi$, 
the proposed regularization Eq.(15) coincides with the one considered by 
Manton and Shifman \cite{mantonshifman}. 

On the other hand, in order to take the light-front 
limit $\theta \rightarrow \pi / 2$, 
one has to be careful for the limiting procedure of taking 
$\epsilon$ and $c \rightarrow 0$. 
As is clear from the regularized chiral charges $Q_{L,R}$, 
one has to take $\epsilon$ first to zero before $c \rightarrow 0$ is taken.
Only in this limit both the left- and the right-chiral charges are 
regularized suitably. 

It is straightforward to evaluate the regularized charges and expand them 
in powers of $\epsilon$
\begin{eqnarray}
\Bigl< Q_L( {\rm reg} ) \Bigr> &=&
{L \over \pi} {1 \over \epsilon} \Big({c \over 1 + s} \Big)
- {e \over \pi} \, L \, A_-^{(o)}
+ {\epsilon \over 2} \Big( {1+ s \over c} \Big)
\Big[ \, {L \over \pi} \, \Big(e \, A_-^{(o)} \Big)^2
- {1 \over 24}{\pi \over L} \Big] + {\cal O}(\epsilon)^2 ,
\nonumber \\
\Bigl< Q_R({\rm reg})  \Bigr> &=&
{L \over \pi} {1 \over \epsilon} \Big( {1 + s \over c} \Big)
 + {e \over \pi} \, L \, A_-^{(o)}
 + {\epsilon \over 2} \Big( {c \over 1+ s} \Big)
 \Big[ \, {L \over \pi} \, \Big(e A_-^{(o)} \Big)^2
- {1 \over 24}{\pi \over L} \Big] + {\cal O}(\epsilon)^2.
\end{eqnarray}

The first terms in Eq.(16) are precisely sources of the infinite constant 
contribution to each charges as the regulator is removed. Therefore we define 
the physical fermion charges by simply dropping them out.
For a constant electric field, $ E = - \partial_+ A_-^{(o)}$, 
we then find that production rates of physical fermion charges are 
$\epsilon \rightarrow 0$
\begin{equation}
{d \over d x^+} \Bigl< Q_L({\rm reg}) \Bigr> = + {e \over \pi} \, L \, E,
\hskip1cm
{d \over d x^+} \Bigl< Q_R({\rm reg}) \Bigr> = - {e \over \pi} \, L \, E .
\end{equation}
Hence,
\begin{equation}
{d \over d x^+} \Bigl< \, Q \, ({\rm reg}) \Bigr> = 0 , \hskip1cm
{d \over d x^+} \Bigl< Q_5({\rm reg}) \Bigr> = {e L \over \pi} \,2 E.
\end{equation}
We see that the electric charge $Q \equiv Q_L + Q_R$ is manifestly conserved, 
consistent with gauge invariance of the proposed regulator.
On the other hand, axial charge $Q_5 \equiv Q_L - Q_R$ is seen anomalous: 
nonzero chiral charge is produced out of the Dirac vacuum at a rate 
$ 2 e \,  L \, E / \pi $. 

This agrees with the results Eq.(12) obtained from Schwinger's 
point-splitting 
method.  More importantly, the current conservation and the axial anomaly in 
Eq.(18) are independent of the interpolating angle $\theta$. In the 
light-front limit, $c \rightarrow 0$, the axial anomaly is correctly reproduced
and remains the same as in the equal-time limit. The crucial point in the 
above regularization is that different regularizations are imposed to the 
left- and the right-moving chiral fermions. See Eq. (\ref{regcharges}). 
The regularization 
depends on the interpolating angle $\theta$. As is clear from the Fig. 1, 
right-moving fermions has to be kept for arbitrarily deep levels inside the 
Dirac sea, while for left-moving fermions a sharp damping is needed. This is 
indeed what happens for the regularized chiral charges in 
Eq.(\ref{regcharges}) and
the way to keep the gauge invariance, hence, charge conservation in a 
manifest way for any interpolation angle
$\theta$ \sl including the light-front limit $\theta = \pi / 2$ \rm. 

The spectral flow of Dirac vacuum fermions also influences the dynamics of the
gauge field by contributing to the total energy $P_+^{(o)}$ of the zero 
mode $A_-^{(o)}$. Let us calculate the contribution coming from the Dirac
vacuum
\begin{equation}
P_+^{(o)} = {1 \over 2 L} \Pi^2_0 + \Bigl < P_{+, ferm} \Bigr>,
\label{totalh}
\end{equation}
where
$P_{+, ferm}$ denotes the energy of fermions in a background of gauge
field
\begin{equation}
P_{+, ferm} = 
\sqrt {1 + s} \, \psi_L^\dagger \, [ i \partial_- - e \, A_-^{(o)}] \, \psi_L 
- \sqrt {1 - s} \, \psi_R^\dagger \, [ i \partial_- - e A_-^{(o)} ] \, \psi_R.
\end{equation}
A short calculation shows
\begin{equation}
\Bigl< P_{+, ferm} \Bigr> =
 \sum_{n \le -1/2}\!\!
\Big({1 + s \over 1 - s}\Big)^{1/2} \, \Big( p_n - e A_-^{(o)} \Big)
- \sum_{n \ge +1/2} \!\!
\Big({1 - s \over 1 + s} \Big)^{1/2} \, \Big( p_n - e A_-^{(o)} \Big).
\end{equation}

Again the sum is not well-defined as it stands because of contributions
from infinitely many modes.
Regularizing the energy in a gauge invariant manner in a way similar to
the chiral charges, we find that
\begin{eqnarray}
\Bigl< P_{+,ferm} \Bigr>  &=&
  \sum_{n \le -1/2} \Big({1 + s \over c}\Big) \,
 \Big( p_n - e  A_-^{(o)} \Big) \,
 \exp \Big[ \epsilon {1 + s \over c} \Big(+ p_n - e  A_-^{(0)} \Big) \Big]
\nonumber \\
 &-& \sum_{n \ge +1/2} \Big( {1 - s \over c} \Big) \,
 \Big(p_n - e \, A_-^{(o)} \Big) \,
 \exp \Big[ \epsilon {1 - s \over c} \Big(- p_n + e \, A_-^{(0)} \Big) \Big]
\nonumber \\
&=& \Biggl[ {d Q \over d \epsilon} \Biggr]_{\epsilon \rightarrow 0}.
\end{eqnarray}

We thus find that
\begin{equation}
\Bigl< P_{+,ferm}  \Bigr>= {e^2 \over \pi c} L \, (A_-^{(o)})^2 ,
\end{equation}
hence, that the total energy Eq.(\ref{totalh}) is given by
\begin{equation}
P_+^{(o)} = {1 \over 2 L} \Pi^2_0 + {e^2 \over \pi c} \, L \, 
\Big(A_-^{(o)} \Big)^2.
\label{totenergy}
\end{equation}
This is precisely the same energy of the zero mode gauge field $A_-^{(o)}$
as was found from the Schwinger's point-splitting method in 
Eq.(\ref{splitanom}).
The second term, which originated from the nontrivial Dirac vacuum deformation
under the gauge field, provides the dynamically generated mass $e /
\sqrt \pi$ of the photon zero mode field $A_+^{(o)}$.  

In this Letter we have shown that both the axial anomaly and the dynamical
mass generation of the two-dimensional massless Schwinger model are correctly 
reproduced from a regularized Dirac vacuum for any interpolating angle 
\sl including the light-front limit \rm $\theta = \pi/2$.
The more standard viewpoint of the anomaly and the dynamical mass generation
was as ultraviolet phenomena. Because of infinitely many quantum states 
involved, the axial anomaly and the dynamical mass generation may alternatively 
be understood as infrared phenomena of the response of the nontrivial
Dirac vacuum under an external gauge field. Naive light-front quantization,
however, loses half of the chiral degrees of freedom, hence, do not admit 
such infrared interpretation. This is the question we have addressed and 
resolved in this Letter. Our results indicate that nontrivial structure of 
the Dirac vacuum persists at the light-front limit so that the correct axial 
anomaly and the dynamical mass generation still arises semiclassically from 
fermion pair production in an external electric field. 
It would be interesting to see if the 
similar interpretation can be made in four-dimensional QCD. Our result would 
also provide further insight to the parton interpretation of the 
spin-dependent deep inelastic scattering \cite{mueller}.
These issue are currently under investigation.

C.-R.J. thanks for the hospitality of the Physics Department and the 
Center for Theoretical Physics at Seoul National University during his 
visit. S.-J.R. thanks S. Brodsky for useful discussions and hospitality
of the Theory Group of SLAC during his visit. 
C.-R.J. was supported in part by U.S. Department of Energy under the 
contract No. DE-FG05-90ER40589.
S.-J.R was supported in part by U.S. NSF-KOSEF Bilateral Grant '94, 
KRF International Coorporation Grant '94, KRF Nondirected Research `94 Grant, 
Ministry of Education BSRI-94-2418 and KOSEF-SRC Program.

\begin{figure}
\caption{
(a) energy-momentum dispersion relation for free fermions at arbitrary 
interpolating angle. Empty circles are vacant states; filled circles are 
occupied states. (b)  energy-momentum dispersion relation at external
electric field. Spectral flows produces vacant negative energy states
and filled positive energy states.}
\label{fig1}
\end{figure}
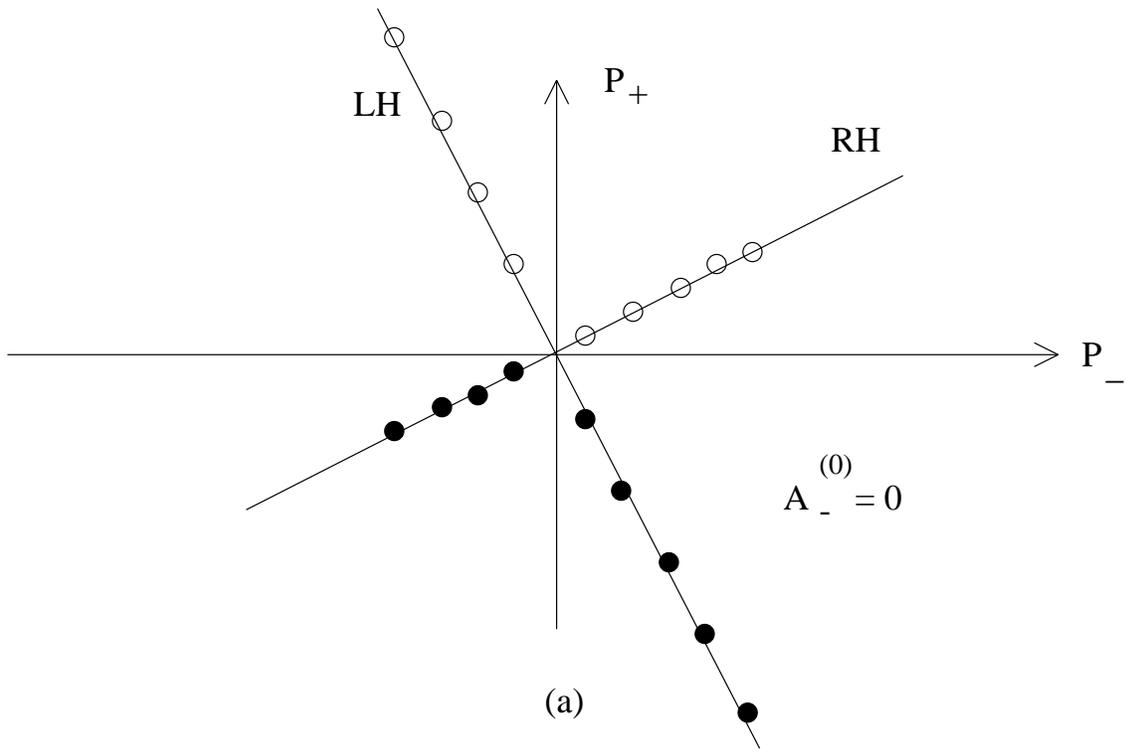
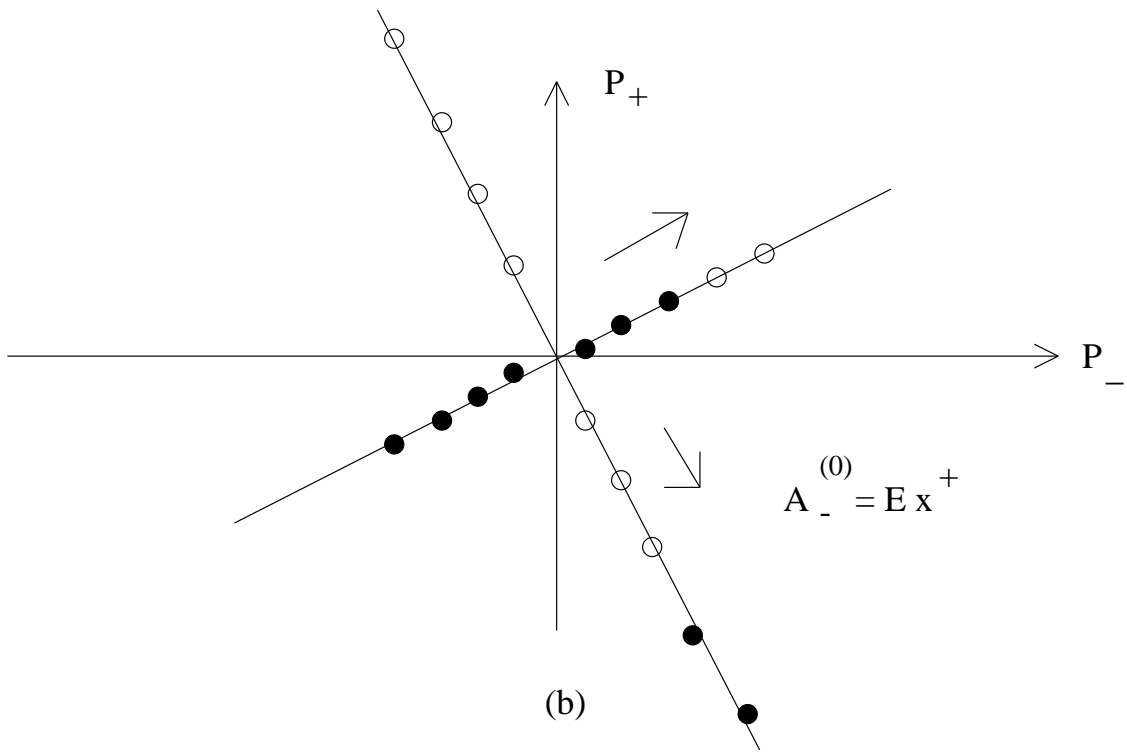
\end{document}